\documentclass[preprint,12pt]{elsarticle}
\pdfoutput=1

\usepackage{amsmath}
\usepackage{amssymb}
\usepackage{graphicx}
\graphicspath{{./}}
\usepackage{booktabs}
\usepackage{tabularx}
\usepackage{longtable}
\usepackage{xcolor}

\definecolor{cLLM}{HTML}{6C5CE7}
\definecolor{cDet}{HTML}{0E9F9C}
\definecolor{cHum}{HTML}{E8963A}
\definecolor{cIn}{HTML}{64748B}
\usepackage{microtype}
\usepackage{url}
\usepackage{hyperref}
\usepackage{orcidlink}
\usepackage{fontawesome5}
\usepackage{tikz}
\usetikzlibrary{shapes.geometric, arrows.meta, positioning, fit, backgrounds}

\usepackage{fvextra}

\journal{}

\begin{document}

\begin{frontmatter}

\title{Towards LLM-Powered Automation of a Dark Matter Constraint Repository}

\author[wustl]{Lanqing Yuan\corref{cor1}\,\orcidlink{0000-0003-0024-8017}}
\ead{yuan.l@wustl.edu}
\cortext[cor1]{Corresponding author.}
\author[wustl]{Karthik Ramanathan\,\orcidlink{0000-0003-4215-7834}}

\affiliation[wustl]{organization={Department of Physics, Washington University in St.\ Louis},
  addressline={One Brookings Drive},
  city={St.\ Louis}, state={MO}, postcode={63130}, country={U.S.A.}}

\begin{abstract}%
Dark matter constraint repositories are critical community infrastructure, yet
the most widely-used are volunteer-maintained, a sustainability risk as results
accelerate. Limits are typically published as figures rather than
machine-readable records, compounding the burden.
We present a large language model (LLM) system that monitors \texttt{arXiv},
extracts limit curves, integrates them as code, and opens pull requests for
human review. Extraction uses a tool-using agent that interprets physical
conventions and checks its results, with a fixed multi-channel pipeline
evaluated as an alternative. We evaluate both approaches on a 329-paper benchmark of measured limits
graded against the upstream-curated repository. An extraction is considered successful when the extracted
and reference mass ranges overlap by more than 90\%, with at least 90\% of
compared values agreeing within a factor of two. For Fable~5, 52.6\% of
eligible curve comparisons meet this criterion with the agent, compared with
20.9\% with the pipeline. Deployed, the pipeline has
generated limit proposals, but none have merged: governance of AI-generated
scientific data is itself unsolved.  {\small\href{https://github.com/FaroutYLq/AutoAxionLimits}{\faGithub}}
\end{abstract}


\end{frontmatter}

\section{Introduction}
\label{sec:intro}

The experimental search for light dark matter is undergoing a
``renaissance''~\cite{snowmass_axion}, with new proposals targeting masses from
$10^{-22}$~eV to several eV~\cite{irastorza_review, kim_review}. The result is an
ever-denser landscape of constraints across at least 15 distinct \emph{coupling
types} (ways dark matter can interact with the Standard Model, e.g.\ axion--photon
or dark-photon kinetic mixing)~\cite{dp_cookbook}, each with its own units,
parameterisations, and physics conventions. Dark matter constraint repositories
compile these bounds into publication-quality \emph{exclusion plots} (the
mass--coupling regions ruled out) used across the field~\cite{axionlimits,
dp_cookbook, ohare_review}. The \texttt{AxionLimits} repository~\cite{axionlimits} tracks
over 600 limits and 160 projected sensitivities, yet is mostly maintained by a single
volunteer who must find new results, digitise figures or tables, convert
parameterisations (e.g.\ the dark-photon kinetic-mixing parameter $\chi$ versus
a coupling constant $g$~\cite{dp_cookbook}), write plotting code,
and regenerate figures. This is community-facing infrastructure with
a single point of failure.

None of these steps is mechanical.
The bound a paper reports is almost never a machine-readable table; it is a
curve on a log--log figure, and often one of several on the same axes (the
paper's new limit drawn alongside older exclusions and projected
sensitivities) that a reader tells apart only from context. Its axes may carry
any of a dozen incompatible parameterisations of the same physics ($g$, $g^2$,
$1/f_a$, $\chi$, $\varepsilon^2$, decay rates, lifetimes), and the conversion
between them frequently lives in the repository's plotting code rather than in
the paper, so the wrong choice is a silent multi-decade error. Reading the
curve then means digitizing a typically log-scaled plot, discarding the projection and
benchmark lines, applying the right dark-matter-density and unit conventions,
and judging whether the result is even a measured bound rather than a forecast.
Each is a small expert judgement, and each done wrong puts the limit orders of
magnitude off, or on the wrong plot entirely. This is why curation has stayed
manual: it is closer to an afternoon of a doctoral student's careful physics
bookkeeping than to a parsing task, and it must be redone for every new paper.

We treat this as a data-processing and software-methods problem rather than an
``LLM demo'': the questions that matter are reproducibility, quantified accuracy,
and auditable provenance. We present a system that partially automates this curation workflow, using a
tool-using agent for extraction and convention conversion, following the
setup described in AxionLimitBench~\cite{yuan2026axionlimitbench}. The agent
chooses how to read each paper, interprets its physical conventions, and can
check and revise the extracted result using information from the text,
figures, and accompanying numerical files. Alternatively, extraction and convention conversion can use a fixed
multi-channel pipeline with deterministic selection rules and a conversion
registry. Both approaches share daily \texttt{arXiv} monitoring with keyword
filtering and LLM relevance classification, followed by automated code
generation, Abstract Syntax Tree (AST)-based insertion, \texttt{Jupyter}
notebook updates, and plot regeneration. Proposed updates are submitted as
pull requests (PRs) with highlighted constraint plots; nothing merges without
a human expert. The system also includes a weekly preprint version
checker.

The task sits in known territory: axion physicists digitise plots today with
\texttt{WebPlotDigitizer}~\cite{rohatgi2024wpd}, and vision--language models
such as \texttt{DePlot}~\cite{liu2023deplot} translate chart images into
tables. The nearest particle physics system is a retrieval-augmented
multi-agent framework for interpreting searches beyond the Standard
Model~\cite{cakir2026limits}, which operates on already-structured
\texttt{HEPData} records; agentic LLM systems have also executed substantial
portions of a collider analysis autonomously~\cite{moreno2026agents}. Our work
sits at the infrastructure end of this spectrum: it starts from the
unstructured paper and grounds the output in an executable, community-facing
repository where every proposed update must survive expert review.
AxionLimitBench~\cite{yuan2026axionlimitbench} describes the extraction
benchmark and agent setup; here we examine their use within the broader
workflow of repository maintenance. To our knowledge, no existing system
automates this full loop for a live scientific repository.

Our contributions are as follows.
(i)~An agent-based approach to extracting limits and interpreting their physical conventions, adopted as the default method.
(ii)~An alternative fixed multi-channel pipeline with validity-first candidate selection, deterministic checks, and standardized convention conversion.
Both approaches feed into automated repository integration and are followed by expert review of the proposed updates.
(iii)~An evaluation on a 329-paper benchmark of measured limits against
curator references~\cite{axionlimitbench_huggingface,yuan2026axionlimitbench},
assessing numerical agreement, mass-range coverage, and repeatability
(section~\ref{sec:eval}).
(iv)~A comparison of three language models using both agent and pipeline
extraction~\cite{axionlimitbench_huggingface}, with paired comparisons between
agent configurations (section~\ref{sec:results}).
(v)~An open-source, human-in-the-loop deployment of the pipeline on the live
literature (section~\ref{sec:integration}).

\section{System architecture}
\label{sec:arch}

\begin{figure}[htbp]
\centering
\includegraphics[width=\textwidth]{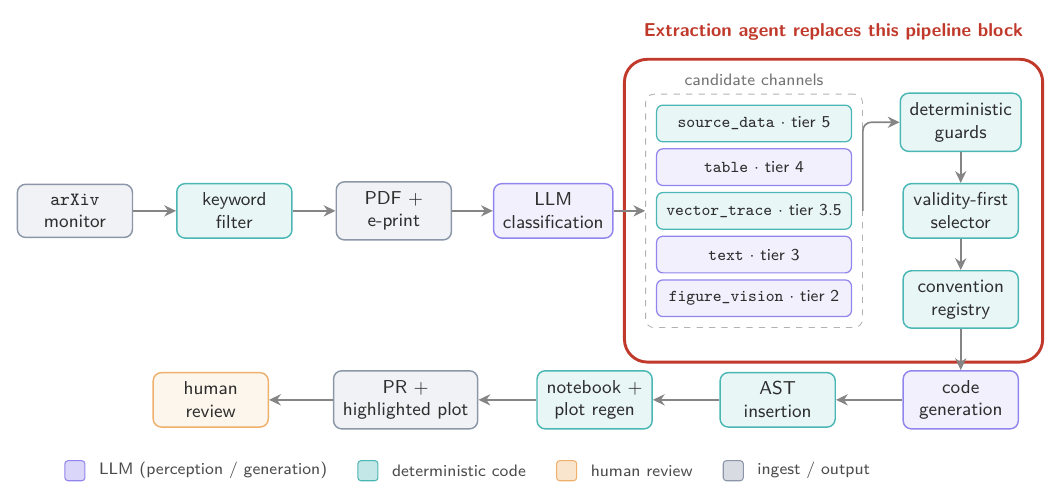}
\caption{Pipeline architecture and its agent-based variant. A paper flows from \texttt{arXiv} (top-left) through
keyword filtering, PDF retrieval, and LLM classification into five competing
extraction channels, a deterministic
decision core (guards, a validity-first selector, and the convention registry),
and code integration, ending at a human-reviewed pull request. In the agent-based workflow, the extraction agent replaces the stages enclosed by the red outline. Colours indicate each stage’s primary role: model-based interpretation (violet), rule-based processing (teal), human review (amber), and input/output handling (slate). Rule-based stages also use model-generated information, so the colours do not imply a strict separation of LLM use and code. Nothing merges without human approval.}
\label{fig:architecture}
\end{figure}

Figure~\ref{fig:architecture} shows the end-to-end workflow, from paper discovery to a human-reviewed pull request. The default approach uses a tool-using agent for extraction and conversion, while the fixed pipeline is retained as a baseline. Both approaches share the discovery, integration, and expert-review stages.

\subsection{Discovery and classification}
A daily \texttt{GitHub} workflow pulls new submissions from the \texttt{arXiv} feed and applies a keyword filter against a curated set of experiment names and coupling jargon. The surviving papers' PDFs are downloaded, and the extraction workflow determines whether each paper reports a new measured or observed exclusion limit and identifies the coupling type. Projected sensitivities are excluded from curation. A sibling workflow reuses the same machinery: a weekly
preprint checker that re-extracts tracked papers when a new \texttt{arXiv} version
appears (it once caught a JWST dark-photon search~\cite{jwst_dp} that demoted a
measured limit to a projection on publication\footnote{The flag and its audit
trail: \url{https://github.com/FaroutYLq/AutoAxionLimits/pull/80}.}) and screens
them for withdrawal from \texttt{arXiv}, proposing the retirement of a limit
whose source paper has been retracted (see the epilogue at the end of this
section). Pipeline state
(processed papers, known preprint versions) is persisted in git alongside the code.

\subsection{Agent-based extraction}
\label{sec:agent-extraction}
We use a tool-using agent to extract limits, following the setup described in AxionLimitBench~\cite{yuan2026axionlimitbench}. Each paper is processed in a separate Claude Code session, with its PDF placed in the working directory and a task prompt appended to the system instructions. The prompt specifies the eligible results, the canonical variable and units for each coupling type, and the required output schema. The agent has seven built-in tools for reading and modifying files, searching, executing shell commands, and retrieving material from the paper's \texttt{arXiv} e-print. Each session has a 30-minute time limit and a \$10 nominal API-cost cap. The full prompt is reproduced in \ref{app:agent-prompt}.

The agent chooses its extraction procedure without a prescribed channel order. It can inspect PDF text, render figures, recover vector paths, or read accompanying numerical files, and write and execute code to calibrate axes and convert quantities using the paper's definitions. It can then check the extracted curve against quoted bounds or other information in the paper and revise its procedure. The final output is a JSON record containing the coupling type, mass--coupling points, declared units and convention, whether the result is a new limit or a projection, a self-reported confidence score, and explanatory notes. Low reported confidence flags an extraction for closer human review. \ref{sec:calib} compares the agent's self-assessment with measured extraction accuracy. The agent performs extraction-time conversions; the separate evaluation procedure is described in section~\ref{sec:eval}.

\subsection{Pipeline baseline}
\label{sec:pipeline-baseline}
The pipeline baseline uses a fixed sequence of model calls and deterministic processing steps, refined using papers from the evaluation sample described in section~\ref{sec:eval}. Its workflow is shown in figure~\ref{fig:architecture}. A bound stated explicitly in the text can provide a direct numerical result without requiring figure digitization. The extraction channels, selection rules, guards, and conversion procedures are detailed in \ref{app:pipeline-extraction}.

\subsection{Integration and review}
\label{sec:integration}
\begin{figure}[htbp]
\centering
\includegraphics[width=0.99\textwidth]{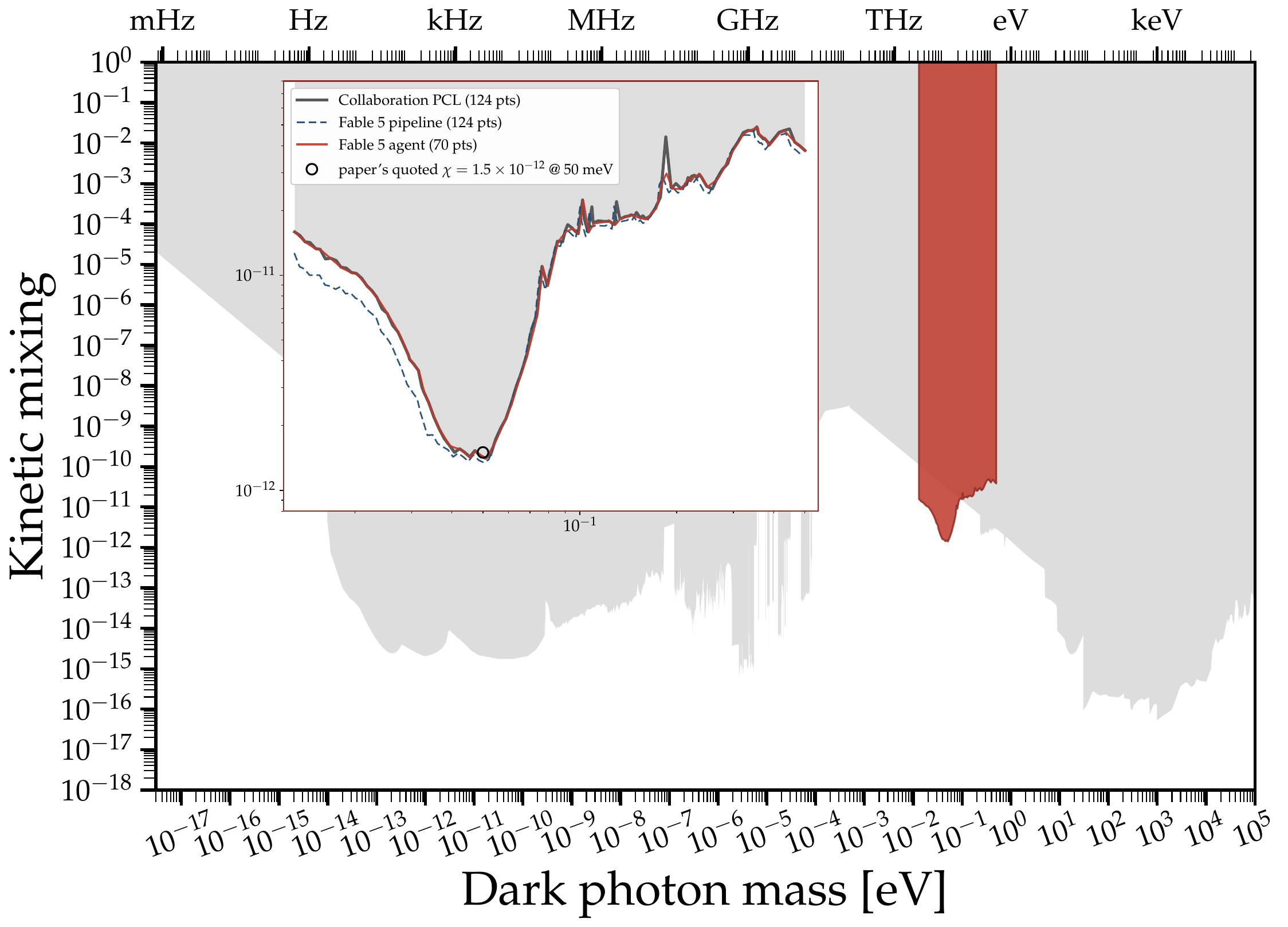}
\caption{The reviewer-facing highlighted plot produced by the case-study run:
the full dark-photon compilation greyed out, with only the proposed
QUALIPHIDE-FIR limit~\cite{qualiphide_fir} in red (13--500~meV, the quoted bound
$\chi=1.5\times10^{-12}$ at 50~meV; the red wedge dips below the existing grey envelope over
13--90~meV, where the paper sets the strongest kinetic-mixing limits to date).
A reviewer can judge the proposal's plausibility without reading
code. The inset zooms into the QUALIPHIDE-FIR band and overlays the Fable~5 agent extraction (red), the pipeline extraction (dashed blue), the paper's quoted power-constrained bound (black circle), and the collaboration's power-constrained limit (grey), all at $\rho_{\rm DM}=0.45$~GeV/cm$^3$.
Over the 122 reference masses common to both comparisons, the median residuals are 0.006~dex for the agent and 0.028~dex for the pipeline.
The agent recovers the power-constrained limit where the pipeline follows the unconstrained curve. This figure demonstrates the curation workflow and should not be cited as a particle-physics result.}
\label{fig:showcase}
\end{figure}

The extracted result is turned into a code change: the LLM
generates a plotting method for the relevant class, and deterministic code
inserts it via Python's AST module (never regex string-matching), adds the
notebook call, and regenerates the figure. Every change becomes
a pull request, but never an automatic merge. Confidence tiers in the PR title
(\texttt{[LOW CONFIDENCE]}, \texttt{[NEEDS REVIEW]}, \texttt{[CONVENTION REVIEW]})
route reviewer attention. The key reviewer aid is a \emph{highlighted plot}:
the full constraint figure rendered with all existing limits greyed out and
only the new proposed curve in red, so a domain expert can assess the
proposal's plausibility without reading code (figure~\ref{fig:showcase}).

For QUALIPHIDE-FIR~\cite{qualiphide_fir}, the paper reports a limit spanning 13--500~meV and quotes $\chi=1.5\times10^{-12}$ at 50~meV. The curves extracted by the agent and pipeline differ from the collaboration's own power-constrained limit by median residuals of 0.006 and 0.028~dex\footnote{Dex denotes a base-10 logarithmic difference: 1~dex corresponds to a factor of ten, and 0.3~dex approximately to a factor of two.}, respectively, evaluated at the same reference masses (inset of figure~\ref{fig:showcase}).\footnote{This comparison uses the exact numerical data behind the paper's figure, without any intermediate digitization by a curator.}

In this example, the pipeline's discrepancy is not simply a tracing error. The collaboration applies a power constraint~\cite{powerconstraint}, raising the observed limit where it falls below the prescribed sensitivity threshold. The pipeline read the plotted curve correctly but did not re-apply this statistical convention, whereas the agent recovers the power-constrained limit. Recognizing this distinction requires interpreting the figure together with the paper's statistical prescription, a task that is difficult to anticipate fully in a fixed extraction procedure.

The example has a postscript that exercises the maintenance half of the system.
On 2026-07-23, two days after the run above, the QUALIPHIDE authors withdrew the
preprint. The weekly checker screens tracked
papers for exactly this. Having detected the retraction the checker proposed the removal in a single pull request for a curator to merge or close.

The pipeline has operated on the live \texttt{arXiv} feed since summer 2026;
every proposal awaits expert disposition as a pull request on our fork, and
deliberately none has been merged into the upstream community compilation. The
human gate is the point of the design rather than a temporary scaffold
(section~\ref{sec:limitations} discusses the continuing need for expert review). Operational details are described in \ref{sec:ops}.

\section{Evaluation methodology}
\label{sec:eval}

\subsection{Ground truth}
\label{sec:gt}
We evaluate on 329 papers spanning the repository's coupling types, chosen as
curated repository data files~\cite{axionlimits} that carry a known \texttt{arXiv} ID
(the full registry holds 437 file entries over 347 distinct \texttt{arXiv} IDs). The benchmark scope is
\emph{measured limits only}: the 25 entries (18 papers) that record
\emph{projected} sensitivities are excluded. The ground truth (GT) for each paper is the
upstream repository's own digitized curve. This choice carries a consequence: some
reference was itself digitized and rescaled from the same paper, so a perfect
extraction still shows a nonzero residual equal to the upstream digitisation
gap. On re-digitized table and text sources that gap is small ($\sim$0.03~dex,
$N=10$), but this is a lower bound: most reference curves were digitized from
figures, where the curator's own digitisation error is plausibly larger, and
the figure-side component has not been measured. The reported differences therefore reflect both extraction errors and an unquantified contribution from uncertainty in the reference data, so differences below 0.1~dex should be interpreted cautiously. 

\subsection{Metrics}
\label{sec:metrics}
We compare in $\log_{10}$--$\log_{10}$ space, using $g$ to denote the coupling variable after both curves have been converted to the same convention. We interpolate the extracted curve in this space and evaluate it at each reference mass $m_i$ within the extracted mass range. If no points can be compared this way, we instead interpolate the reference curve at extracted masses lying within the reference range. The residuals from the two comparisons are not combined. The difference at each point is
\begin{equation}
  r_i = \bigl|\log_{10} g_{\rm ext}(m_i) - \log_{10} g_{\rm truth}(m_i)\bigr|
  \label{eq:residual}
\end{equation}
where $g_{\rm ext}$ and $g_{\rm truth}$ are the extracted and reference coupling values, respectively. These differences are measured in dex; $0.3$~dex corresponds approximately to a factor of two. A paper's \emph{residual} is the median of $r_i$ over the compared mass points. When a paper has multiple eligible references in a compatible coupling plane and convention, we select the one with the smallest finite median residual and retain it for all metrics. Both are described in \ref{app:selection}.

Pointwise agreement is the fraction of compared points within 0.3~dex. We report the percentage of papers for which this fraction is at least 90\%.

For mass-range coverage, let $R$ and $E$ denote the mass ranges covered by the reference and extracted curves, respectively, excluding any explicitly recorded gaps, and define
\begin{equation}
  C_{\rm ref}=\frac{|R\cap E|}{|R|},
  \qquad
  C_{\rm ext}=\frac{|R\cap E|}{|E|},
  \label{eq:coverage}
\end{equation}
where $|\cdot|$ measures the total length of these ranges in $\log_{10}m$.
Reference coverage measures how much of the reference range is recovered, while extraction coverage measures how much of the extracted range lies within the reference.
Joint success requires both coverages $>90\%$ and at least 90\% of compared points within 0.3~dex, for the same paper and reference.
Coverage weights log-mass intervals uniformly, whereas pointwise agreement retains the scorer's discrete comparison points.

A paper enters the numerical comparison when its extraction can be compared
with a reference curve for the same coupling type and yields a finite residual.
Papers whose extracted and reference quantities cannot be converted to compatible
conventions are excluded. These exclusions occurred only for the pipeline in
the evaluated runs; all evaluation outcomes are reported in \ref{app:selection}.

Residual and pointwise-agreement statistics use all finite comparisons, with other evaluation outcomes reported separately in \ref{app:selection}.
For coverage and joint success, we exclude nine point-only references (limits specified at a single mass) per configuration and the documented zero-mass representation cases\footnote{These cases contain mass-zero entries that prevent an unambiguous comparison of logarithmic mass ranges. All three exclusion counts refer to the pipeline configurations.}: three for the Fable pipeline, two for Opus, and five for Haiku.
An extraction giving a limit at only one mass remains included when its reference covers a mass range and fails the coverage requirement.
Table~\ref{tab:results} uses all $N$ finite comparisons for pointwise agreement and the remaining $N_{\rm curve}$ comparisons for coverage and joint success. These percentages are conditional on a comparison rather than end-to-end success rates.
For the pointwise-agreement and joint-success rates, we estimate 95\% confidence intervals by resampling papers with replacement 20,000 times and taking the 2.5th and 97.5th percentiles of the resulting percentages. These intervals describe variation across papers, not between repeated extraction runs.

\subsection{Confound control}
\label{sec:confound}
Three factors affect how the reported accuracy should be interpreted:
differences in model and extraction workflow, variation between repeated
runs, and overlap between the papers used for development and evaluation.
We compare models on shared papers, measure run-to-run variation, and
document the overlap.

\emph{Model comparison.} We compare agent configurations on papers with
finite residuals in both runs (section~\ref{sec:modeleffect}). Pairing holds
the evaluated papers fixed, but does not isolate model capability from
differences in agent configuration.

\emph{Run-to-run stochasticity.} Because the extraction agent samples from an
LLM, a single benchmark run is a noisy estimate of accuracy, so run-to-run
variation is measured directly. A frozen random 100-paper subset is
processed twice with the same Fable~5 agent configuration, yielding finite
residuals in both runs for 92 papers. On these shared papers, the aggregate
median residuals are 0.028 and 0.039~dex, while the median absolute change in
per-paper residual is 0.002~dex. Accuracy scores are therefore closely
reproducible for most papers, but occasional larger changes remain: seven
papers differ by at least 0.1~dex. The two largest changes arise from
different physical-scenario and rescaling choices, illustrating variation
beyond curve tracing; the full distribution and outcome transitions are
detailed in \ref{sec:repeatability}.

\emph{Agent--pipeline comparison.} The pipeline uses fixed extraction,
selection, and convention rules, some informed by failures on the benchmark
papers. The agent instead chooses its extraction steps and interprets the
paper's conventions using the task prompt reproduced in \ref{app:agent-prompt}, without
using those pipeline rules. The comparison therefore evaluates the two
workflows as a whole, rather than isolating the effect of any individual rule.

\section{Results}
\label{sec:results}

\subsection{Headline benchmark results}
\label{sec:headline}
\begin{figure}[htbp]
\centering
\includegraphics[width=0.61875\textwidth]{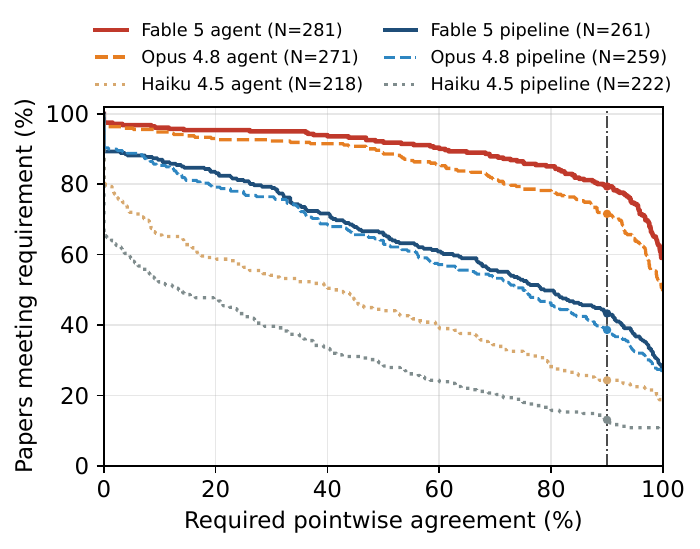}
\caption{Percentage of compared papers meeting a given pointwise-agreement
requirement for the six benchmark configurations. Pointwise agreement is the
fraction of compared points within 0.3~dex of the reference. The vertical line
marks the 90\% requirement used in Table~\ref{tab:results}; the dots give the
corresponding passing percentages. Each curve uses all finite comparisons,
matching the pointwise-agreement block of the table; coverage is not required.}
\label{fig:cdf}
\end{figure}

Table~\ref{tab:results} compares three models using either an autonomous extraction agent or the fixed pipeline: Claude Fable~5~\cite{anthropic_claude}, Claude Opus~4.8, and the smaller Claude Haiku~4.5. With one agent session per paper, the Fable~5 agent yields finite comparisons for 281 of the 329 processed papers; the remaining outcomes are listed in Table~\ref{tab:disposition}. Of the 272 comparisons eligible for curve evaluation, 52.6\% meet the joint coverage and pointwise-agreement criterion (95\% confidence interval: 46.7--58.5\%). The corresponding pipeline result is 20.9\% (16.1--26.1\%) among 249 eligible comparisons. Both coverage requirements are met by 64.0\% of agent comparisons and 54.6\% of pipeline comparisons. Among all finite comparisons, 79.4\% of the 281 agent results and 43.3\% of the 261 pipeline results pass the pointwise-agreement requirement. Figure~\ref{fig:cdf} shows how the passing fraction varies with the required pointwise agreement; the distributions of both coverage measures are shown in figure~\ref{fig:coverage-distributions}.

Beyond the curve itself, the accuracy of both workflows' categorical outputs
is also measured. Coupling-type classification is 99.7\% accurate for the
Fable~5 agent and 98.3\% for its pipeline counterpart, each evaluated on 291
papers against the benchmark's accepted coupling types (grading rules in
\ref{app:selection}).

\begin{table}[tp]
\caption{Comparison of three language models (Fable~5, Opus~4.8, and
Haiku~4.5), each using either an autonomous extraction agent or the fixed
multi-channel pipeline, on a common benchmark of 329 measured-limit papers.
Pointwise agreement uses all $N$ finite comparisons; coverage and joint success
use the $N_{\rm curve}$ comparisons remaining after the exclusions described
in the text. Pointwise agreement $\geq90\%$ requires at least
90\% of compared points to lie within 0.3~dex of the reference. Joint success
requires this condition and both mass-range coverages $>90\%$, using the same
reference. Brackets give 95\% bootstrap confidence intervals over papers.}
\label{tab:results}
\centering
\begingroup
\fontsize{9}{10.8}\selectfont
\setlength{\tabcolsep}{2pt}
\renewcommand{\arraystretch}{1.16}
\begin{tabularx}{\textwidth}{@{}p{108pt}*{6}{>{\centering\arraybackslash}X}@{}}
\toprule
 & \multicolumn{3}{c}{\textbf{Agent}} & \multicolumn{3}{c}{\textbf{Pipeline}} \\
\cmidrule(lr){2-4}\cmidrule(l){5-7}
 & Fable 5 & Opus 4.8 & Haiku 4.5 & Fable 5 & Opus 4.8 & Haiku 4.5 \\
\midrule
\multicolumn{7}{l}{\textit{Pointwise agreement}} \\
Finite comparisons, $N$ & 281 & 271 & 218 & 261 & 259 & 222 \\
\shortstack[l]{Pointwise agreement\\$\geq90\%$ (\%)} & \shortstack{79.4\\{\fontsize{7}{8}\selectfont [74.4--84.0]}} & \shortstack{71.6\\{\fontsize{7}{8}\selectfont [66.1--76.8]}} & \shortstack{24.3\\{\fontsize{7}{8}\selectfont [18.8--30.3]}} & \shortstack{43.3\\{\fontsize{7}{8}\selectfont [37.2--49.4]}} & \shortstack{38.6\\{\fontsize{7}{8}\selectfont [32.8--44.8]}} & \shortstack{13.1\\{\fontsize{7}{8}\selectfont [9.0--17.6]}} \\
\midrule
\multicolumn{7}{l}{\textit{Coverage and joint success}} \\
Eligible curves, $N_{\rm curve}$ & 272 & 262 & 209 & 249 & 248 & 208 \\
$C_{\rm ref}>90\%$ (\%) & 74.3 & 69.1 & 48.8 & 65.9 & 58.5 & 56.2 \\
$C_{\rm ext}>90\%$ (\%) & 85.3 & 87.0 & 68.9 & 81.5 & 78.2 & 56.2 \\
Both coverages $>90\%$ (\%) & 64.0 & 61.5 & 31.1 & 54.6 & 47.6 & 26.9 \\
\textbf{Joint success (\%)} & \shortstack{\textbf{52.6}\\{\fontsize{7}{8}\selectfont [46.7--58.5]}} & \shortstack{\textbf{45.4}\\{\fontsize{7}{8}\selectfont [39.3--51.5]}} & \shortstack{\textbf{8.6}\\{\fontsize{7}{8}\selectfont [5.3--12.4]}} & \shortstack{\textbf{20.9}\\{\fontsize{7}{8}\selectfont [16.1--26.1]}} & \shortstack{\textbf{16.5}\\{\fontsize{7}{8}\selectfont [12.1--21.4]}} & \shortstack{\textbf{3.4}\\{\fontsize{7}{8}\selectfont [1.0--5.8]}} \\
\bottomrule
\end{tabularx}
\endgroup
\end{table}

\subsection{Why the agent improves extraction}
\label{sec:agent-improvement}
On the 248 papers eligible for coverage evaluation in both Fable~5 workflows, the
agent meets the joint coverage and accuracy criterion in 54.8\% of cases,
compared with 21.0\% for the pipeline. The saved extraction records
illustrate how the agent's ability to adapt its procedure to each paper
helps recover curve details and carry physical interpretations through to
the numerical output.

The agent can adapt how it reads a figure to the structure of the plotted
result. For the RADES axion search~\cite{ahyoune2025rades}, whose exclusion
curve spans a very narrow mass range, the pipeline returns a few
approximate values from the text, while the agent extracts the detailed
curve from the figure. The median residual falls from 0.145 to 0.020~dex,
with both extractions already covering more than 99\% of the reference
mass range.

The agent can also use the paper's physical relations to interpret and
convert the extracted result. For the DESI decay search~\cite{wang2023desi},
it converts the plotted decay rate into a photon coupling using the
paper's equation and the required unit conversions; the pipeline states
the equation but leaves the values unconverted. In a figure containing
both a theoretical lifetime bound and an observational X-ray
exclusion~\cite{ferreira2022electron}, the pipeline returns the lifetime
bound described in the text; the agent uses its characteristic mass
dependence to distinguish the two similarly shaded regions and extract
the observational limit. Physical interpretation and iterative checks
against the paper allow the agent to revise its extraction in ways that
are difficult to anticipate and encode in fixed rules.

\subsection{Model comparison}
\label{sec:modeleffect}
We compare the three agent configurations by pairing papers with finite
residuals in both configurations. Comparisons across and within model
capability tiers show different patterns.

\emph{Across capability tiers}, the difference is large. Among the 218 papers
with finite comparisons for both agents, the median per-paper residual
difference is Haiku $-$ Fable~5 $= +0.277$~dex. Fable~5 has a lower residual
by more than 0.05~dex on 166 papers, Haiku on four, and the remaining 48
differ by at most 0.05~dex. Among the 209 papers eligible for coverage
evaluation in both configurations, joint success is 54.1\% for Fable~5
and 8.6\% for Haiku.

\emph{Within a capability tier}, the typical difference is much smaller.
Among the 270 papers with finite comparisons for both agents, the median
per-paper residual difference is Fable~5 $-$ Opus~4.8 $= -0.001$~dex.
Fable~5 has a lower residual by more than 0.05~dex on 49 papers, Opus on 14,
and the remaining 207 differ by at most 0.05~dex. Nevertheless, among the
261 papers eligible for coverage evaluation in both configurations, joint
success is 52.9\% for Fable~5 and 45.2\% for Opus. Thus, similar residuals on
most papers do not imply equal rates of recovering a sufficiently complete
and accurate curve. Computational costs are discussed in \ref{sec:cost}.

\subsection{Remaining challenges in extraction and evaluation}
\label{sec:channelvalue}
\begin{figure}[htbp]
\centering
\includegraphics[width=0.9\textwidth]{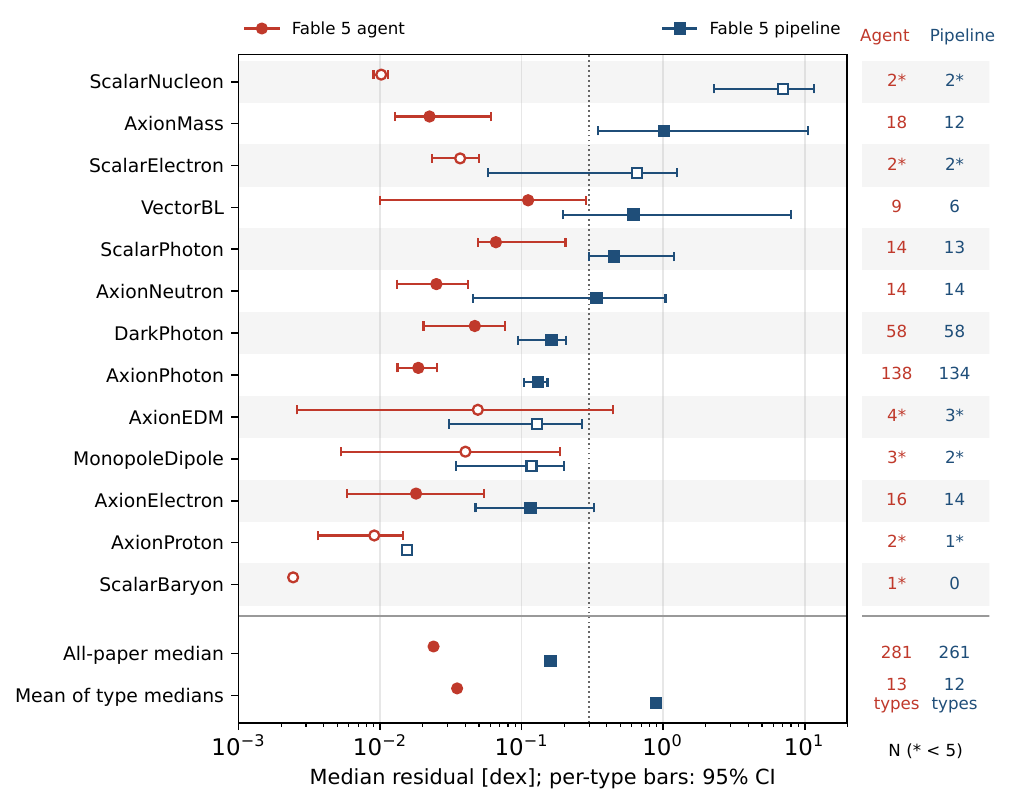}
\caption{Median residual by coupling type with bootstrap 95\% CIs for the
Fable~5 agent and pipeline. The overall summaries are the median residual across papers (micro-median) and the mean of the median residuals for individual coupling types (macro average). The latter gives each coupling type equal weight. The pipeline's micro--macro gap is concentrated in
rare, convention-heavy types. Small-sample types ($N<5$) use hollow markers;
sample counts are shown at right.}
\label{fig:pertype}
\end{figure}

The low median residuals in Figure~\ref{fig:pertype} do not guarantee agreement along every curve. Some discrepancies arise from choosing between results published under different physical assumptions. In the Axinovae study~\cite{fox2023axinovae}, for example, the agent selects the optimized bound, whereas the reference contains the conservative bound. Their disagreement therefore does not by itself indicate incorrect tracing. This illustrates why numerical agreement must be assessed alongside the identity of the selected result.

Preserving curve structure presents a separate challenge. In the inset of Figure~\ref{fig:showcase}, for example, the agent follows the overall limit closely but misses a narrow peak in the collaboration's curve. Other discrepancies arise during evaluation: the MADMAX~\cite{madmax2025prototype} and ADMX Sidecar~\cite{boutan2018sidecar} outputs identify separate mass windows in their notes, but the numerical comparison connects points across those gaps. Improving fidelity therefore involves both extracting the curve accurately and preserving its structure through evaluation.

The reference compilation~\cite{axionlimits} can also differ in scope from the supplied paper. Some entries bring together results from multiple publications~\cite{linden2025tridents,nguyen2026decay} or continue a flat bound beyond the plotted range~\cite{ning2024m82}; others cover a narrower mass interval than the published figure~\cite{benabou2025polarization}. Such differences need not indicate errors in either the extraction or the compilation. The two coverage measures therefore quantify agreement with the stored mass intervals, which may not directly reflect completeness relative to the supplied paper.

\section{Limitations and future work}
\label{sec:limitations}
Agreement with a reference curve does not by itself establish that the intended physical result has been recovered. Expert review must still check the selected scenario, assumptions, and conventions against the paper. Further development of the agent harness\footnote{The software that manages the agent's tools and execution.} may help by improving how these choices are recorded and checked, together with output formats that preserve separate mass windows and curve structure. Improvements in frontier-model capabilities may also help overcome the remaining difficulties in physical interpretation and curve extraction. Evaluating these developments remains future work.

The system provides an initial extraction for expert review, but the resulting reduction in curation effort has not been measured. The Fable~5 agent has a median estimated extraction cost of \$1.48 per paper at nominal API rates~\cite{yuan2026axionlimitbench}. Reviewing and correcting proposals may still require substantial work, including manual extraction when necessary. The computational cost is expected to be far below the cost of the expert time saved, but this expectation remains untested by a timed comparison with manual curation.

Publishing machine-readable limits would reduce the need for extraction from figures. Infrastructure such as HEPData~\cite{hepdata} and the recommendations of the LHC Reinterpretation Forum~\cite{reinterpretation_forum} already support reusable result publication. Wider adoption for light-dark-matter limits, with explicit conventions and physical assumptions, would simplify both automated extraction and expert verification.

\section{Conclusions}
\label{sec:conclusion}
We developed a system that monitors the dark-matter limit literature, extracts curves, and prepares repository updates for expert review. The revised workflow uses a tool-using agent for extraction, with the fixed pipeline retained as a baseline. On the 329-paper benchmark~\cite{axionlimitbench_huggingface,yuan2026axionlimitbench}, 52.6\% of Fable~5 agent extractions eligible for curve evaluation meet the joint coverage and pointwise-agreement criterion, compared with 20.9\% for the pipeline. The saved extraction records help explain the agent's advantage: it adapts its figure reading, interprets physical conventions, and checks its results against the paper. Expert review remains necessary to verify the selected result and its assumptions, and the reduction in curation effort is expected but has not yet been measured. We offer the system, benchmark, and supporting analyses as tools for maintaining limit compilations for the dark-matter community.

\clearpage
\appendix
\makeatletter
\@addtoreset{figure}{section}
\@addtoreset{table}{section}
\makeatother
\section{Pipeline methods and operation}
\label{app:pipeline-extraction}

\subsection{Multi-channel extraction}
\label{sec:channels}
An extraction \emph{channel} is a method for obtaining limit values from a paper’s text, tables, figures, or accompanying data files. 
Extraction turns a paper into a structured record: coupling type, a list of
$(\text{mass}, \text{coupling})$ data points, the declared units/convention,
whether the result is a new limit or a projection, an extraction-confidence
score, and notes. The confidence is the model's own $[0,1]$ self-assessment of
the extraction, not a statistical quantity. It serves as a low-priority
tie-breaker in candidate selection (\ref{sec:select}) and flags
low-confidence extractions for human review in the PR title (section~\ref{sec:integration}).
These uses rely on confidence ranking extraction quality rather than representing a calibrated probability. Whether it ranks in
practice is tested in \ref{sec:calib}. Rather than one extraction path, the pipeline produces
\emph{candidates} through up to five channels of differing semantic trust
(table~\ref{tab:channels}), staged as: a cheap particle physics interaction type pre-classification pass; a text/table read of the parsed PDF (\texttt{PyMuPDF}~\cite{pymupdf}); a cheap
axis-calibration pass that identifies the relevant figure and reads its axis
labels and ranges; and, when needed, a full vision trace of the rendered figure pages. Two
channels are deterministic: \texttt{source\_data} harvests numeric files
shipped in the paper's own e-print source (for example, the \texttt{.dat} tables behind
\texttt{pgfplots} figures, or separately uploaded ancillary data files), exact where present, and \texttt{vector\_trace} extracts
curve geometry directly from the vector paths of PDF figures, with a single
cheap LLM call used only to identify which path is the limit curve.
As a defence against prompt injection, the paper text is stripped of control
characters and enclosed in delimiters, with instructions to treat the enclosed text as data.

The cheap pre-classification tagged QUALIPHIDE-FIR as
\texttt{DarkPhoton} at confidence 0.97; the axis-calibration pass then located
the limit figure shipped in the e-print source
(\texttt{qualiphide\_dp\_limit.pdf}). Because that figure is a vector graphic,
the deterministic \texttt{vector\_trace} channel could extract the curve
geometry directly from its PDF paths: among three high-confidence path
candidates the single cheap identification call picked ``the dark solid
observed-limit curve spanning 13--487~meV,'' yielding a 124-point candidate at
confidence 0.85. The text channel independently contributed the paper's only
numeric bound (the power-constrained
$\chi=1.5\times10^{-12}$ at 50~meV) as a single-point candidate at
confidence 0.70. With a deterministic trace and a corroborating text read in
hand, the expensive vision channel was skipped entirely.

\begin{table}[htbp]
\caption{Candidate extraction channels and their semantic trust tiers. Higher tiers are defined with less reliance on model interpretation: supplied numerical data require none, parameter bounds stated in the text require reading comprehension, and visually tracing a plot requires the most interpretation. The channel selector considers tier only after all validity criteria, preferring higher-tier channels among equally valid candidates (\ref{sec:select}).}
\label{tab:channels}
\centering
\small
\begin{tabularx}{\textwidth}{@{}lcX@{}}
\toprule
\textbf{Channel} & \textbf{Tier} & \textbf{Nature} \\
\midrule
\texttt{source\_data}  & 5   & deterministic numeric files from the paper's own e-print source \\
\texttt{table}         & 4   & LLM read of a tabulated bound \\
\texttt{vector\_trace} & 3.5 & deterministic vector-path geometry; curve identity by one cheap LLM call \\
\texttt{text}          & 3   & LLM read of bounds stated in prose \\
\texttt{figure\_vision}& 2   & LLM trace of the rendered exclusion plot \\
\bottomrule
\end{tabularx}
\end{table}

\subsection{Validity-first candidate selection}
\label{sec:select}
Each candidate is scored by a \emph{lexicographic quality tuple}: The selector compares candidates using a fixed sequence of checks. The first check on which two candidates differ determines which is preferred. Physical validity and interpretable units and conventions take priority. Among candidates that match on these checks, the selector considers channel tier, agreement with an independent channel, the model’s reported confidence, and finally the number of data points, in that order. Text and table candidates containing at most three points receive tier 1. Higher confidence or more points cannot outweigh a validity problem. The criteria are listed below.

\emph{(1) Values in physically allowed ranges.} Each coupling type has fixed,
broad ranges for mass (in eV) and coupling (in the standard convention), based
on its known physical range and checked against the repository's data. A
candidate passes when its median mass and coupling lie within these ranges.
This check catches large unit errors, such as a mass read in $\mu$eV but
labelled eV or a missing power-of-ten factor in a coupling. It does not assess
the paper's underlying physics.

\emph{(2) Curve shape and mass range.} A nearly flat trace\footnote{The
extracted coupling varies by at most 0.05~dex.} or one covering too little of
the figure's mass range receives a lower selection priority; this check does
not by itself discard the candidate. The minimum span is half the figure's
logarithmic mass range, capped at one decade; when the figure's range is
unknown, one decade is used. Text and table extractions are exempt. These
are heuristic preferences: legitimate flat or narrow limits can also be
penalized.

\emph{(3) Correctable by a known unit factor.} If the median coupling lies
outside the allowed range, the selector checks whether multiplying by a
factor from a fixed list of unit-prefix powers of ten brings it back inside
(for example, restoring a missing $10^{-14}$). Among candidates tied on the
earlier checks, a correctable reading ranks above one that remains outside
the range. The factor must come from this list; it is not fitted to the data.

\emph{(4) Supported convention.} The candidate's declared convention must
match a reviewed entry in the conversion registry (\ref{app:conventions}),
allowing code to convert the values to the repository's canonical variable.

\emph{(5) Channel tier.} Higher tiers indicate less reliance on model
interpretation (table~\ref{tab:channels}). Tier is considered only when
candidates match on all four validity checks.

\emph{(6) Agreement with an independent channel.} A candidate is preferred
if another channel produces a comparable extraction agreeing within a factor
of three (a ratio in $[1/3,3]$, or about 0.48~dex). This preference applies
only when candidates tie on the preceding criteria. It is separate from the
rejection check in \ref{sec:guards}: a curve extracted visually or
from PDF vector paths is rejected if it differs from an in-range text bound
by more than a factor of 100 (2~dex) where their mass ranges overlap.

\emph{(7) Self-reported confidence.} The model's confidence score, between
0 and 1, breaks ties remaining after the preceding checks. Whether higher
confidence corresponds to better extractions is examined in \ref{sec:calib}.

\emph{(8) Point count.} If candidates tie on every preceding criterion, the
one with more extracted points is preferred. This is the only role of point
count in selection.

For QUALIPHIDE-FIR the deterministic \emph{vector trace} won over the text
channel's stated bound: geometry the paper itself ships sits above a
single stated number on the tier order (table~\ref{tab:channels}), and the
selector's recorded quality tuple, which is valid on all four criteria, tier 3.5,
confidence 0.85, 124 points, carried it.

\subsection{Deterministic guards}
\label{sec:guards}
Before selection, automatic checks reject unsuitable candidates or lower their priority based on the model’s notes, the mass range stated in the abstract, and the relevant physical regime. For example, a candidate may be rejected if its mass range lies far outside the range stated in the abstract. Curves extracted visually or from PDF vector paths are also rejected if their median absolute logarithmic difference from an in-range text extraction exceeds 2~dex, provided at least two text points overlap the curve\footnote{This rejection check is distinct from the preference for agreement between channels used during selection (see \ref{sec:select})}. Finally, automatic correction of unit-prefix errors, such as confusing GeV with eV, is disabled when the declared convention is flagged for review. This reduces the risk of a valid squared-coupling ($g^2$) axis being mistaken for a unit error and incorrectly rescaled.

For QUALIPHIDE-FIR, at 50~meV the vector trace reads
$\chi\simeq1.35\times10^{-12}$, close to the text channel's stated
$1.5\times10^{-12}$. The single text point was insufficient to activate this
cross-check, and the vector trace was selected by the ranking described above.

\paragraph{Pipeline guard outcomes}
In the Fable~5 pipeline benchmark run, cross-checks against values stated in the text rejected 26 candidates extracted from figures, while unit and convention checks flagged 20 papers for human review. These 20 papers were flagged during extraction. This count differs from the unresolved-convention category in table~\ref{tab:disposition}, which records cases where an unsupported convention prevents comparison with the reference curve during evaluation. These counts show how often the checks acted, but not how often they rejected a valid extraction or lowered its selection priority.

\subsection{Convention canonicalization}
\label{sec:conv}
\label{app:conventions}
The same physical bound can be expressed using different vertical-axis quantities across papers, for example including $g$, $g^2$, $g^2/4\pi$, $1/f_a$, $\Lambda$, $\varepsilon^2$, decay rates, and lifetimes in axion. The repository preserves this variety: even within a single coupling type, data files store different quantities, which are converted to the plotted coupling only by the plotting code. Comparing these curves without accounting for those differences can make correct extractions appear wrong by several orders of magnitude. We prevent this through three rules, detailed below: each extraction channel must accurately identify the quantity and convention of its output; the vision channel must return values exactly as plotted, without converting them; and all pipeline conversions must use one reviewed set of conversion definitions. The conversion factors were derived with the agentic physics assistant Get Physics Done (GPD)~\cite{gpd}, rather than simply requested from a model without harness, then checked in code against the repository’s plotting source and verified against the cited primary literature~\cite{damour_donoghue_dilaton, diluzio_landscape}.

Convention canonicalization converts quantities to the repository's standard
variable and convention. We refer to this variable as the \emph{canonical
variable}.

\emph{1. Identify the output quantity.} Every channel must accurately state the
quantity and convention of the values it returns, so later processing does
not have to infer them from their magnitudes.

\emph{2. Preserve the plotted values.} The vision channel reads values without
converting them and identifies the plotted quantity. If its declaration
claims the standard convention but contradicts the independently read axis
label, the axis label takes precedence. Allowing the model to convert values
previously led to conversions being applied twice or skipped entirely,
producing errors of about 8~dex in otherwise correct readings
(\ref{sec:anatomy}).

\emph{3. Manage conversions in one place.} All pipeline conversions use one
reviewed conversion registry.

\paragraph{Handling unsupported quantities and conventions}
The pipeline requires human review when a declared quantity or convention
is not supported by the reviewed conversion definitions. It checks the full
declaration rather than looking for familiar symbols within it: a declaration
may contain a known coupling symbol while describing a different quantity,
such as a decay width. Partial text matching previously allowed such cases
to pass without a review flag. A separate check identifies incorrect powers
of units, such as $\varepsilon$ in GeV$^{-1}$ or $g_{a\gamma}$ in GeV$^{-2}$;
these require review unless an approved conversion exists for that exact
variable. Candidates that cannot be converted receive a
\texttt{[CONVENTION REVIEW]} flag and a capped confidence score. They are sent for human review. During evaluation, cases identified by the scorer as convention mismatches are excluded from numerical scoring.
Values declared as already ``converted from \dots'' are not
converted again.

\paragraph{Mass-dependent conversion}
The conversion registry's first mass-dependent converter relates the oscillating
electric dipole moment $d_n$ [e$\cdot$cm] to the operator coupling through
the dark-matter field amplitude:
$g_{an\gamma}=d_n\,m_a/\sqrt{2\rho_{\rm DM}}$.
This change of variable has no free parameters. It was derived with GPD and
spot-checked against papers that publish curves in both variables, giving
agreement of 0.13--0.26~dex.

\paragraph{Per-converter magnitude guards}
Each converter checks whether the input values are plausible for the declared
quantity before applying a conversion. How the pipeline resolves inconsistent
declarations, and the effect of these checks on accuracy, are described in
\ref{sec:anatomy}.

\paragraph{Axis read-back as a classification cross-check}
The \emph{axis read-back} is the reading of axis labels during the initial
figure inspection. These labels provide a second
check on the coupling type. If they unambiguously identify another coupling
within a group of easily confused types, the pipeline updates the type.
Examples include $g_{B-L}$ versus $\varepsilon$, $d_g$ versus $d_e$, and
e$\cdot$cm versus GeV$^{-2}$. Contradictions between different coupling
families are flagged for review without changing the type. Unreadable axes
leave the existing classification unchanged and do not stop processing.
Axes showing combinations of symbols, or $\varepsilon$ alone, are treated
as ambiguous because $B\!-\!L$ papers also use that symbol. These restrictions
were introduced after examining benchmark errors.

Table~\ref{tab:conventions} catalogues representative conversion families. The registry's use in benchmark scoring and the handling of unusual reference files are described in \ref{app:benchmark-conventions}. Treating these factors as a small but exacting derivation task,
rather than a prompting problem, is what made the registry trustworthy enough
to apply automatically.

QUALIPHIDE-FIR needed no registry conversion: its declared plotted quantity,
kinetic mixing $\chi$, is already the repository's canonical variable, and its
declared dark-matter density $\rho=0.45$~GeV/cm$^3$ is the repository's own
convention, so both the conversion and the plot-time density rescale
($\chi\propto1/\sqrt{\rho}$, applied by the plotting method) ran
at exactly identity.

\subsection{Operational safeguards}

\label{sec:ops}
Four safeguards support routine operation of the pipeline baseline:

\emph{Stopping when the service is unavailable.} Billing and authentication
errors stop the run without marking the current paper as processed or failed.
The pipeline does not substitute a default answer, which would make a service
failure appear to be a result about the paper.

\emph{Checking which model is running.} At startup, the pipeline checks the
identity of the model actually in use. This check was added after an unnoticed
model substitution affected a benchmark.

\emph{Controlling costs.} When a rate limit is reached, model calls are retried
with progressively longer delays. Direct extraction from numerical files and
vector paths limits model use where possible. Per-paper computational costs are summarized in
\ref{sec:cost}. Computational costs are expected to be far below the cost of the expert time saved, but the time savings have not been measured.

\emph{Transport: API or subscription access.} The transport is the method
used to access the model. One configuration setting selects
between an API billed by token use and a local Claude Code subscription.
The subscription option runs the same pipeline code through the \texttt{claude}
command-line tool, using authentication stored in the operating-system
keychain without an API key. This gives volunteer maintainers a flat-rate
option for the daily workflow. For both access methods, a one-token test
request checks authentication and available usage before extraction begins.

\section{Lessons from pipeline development}
\label{sec:discussion}
This appendix collects lessons from developing and evaluating the pipeline baseline. It closes with two examples of
conversion errors and the checks that prevent them (\ref{sec:anatomy}).
These lessons can also guide other literature-extraction pipelines.

\subsection{What helped}
\begin{itemize}
\item \emph{Validity-first candidate selection.} Extraction channel candidates are ranked
  by validity before any other preference. An incorrect trace cannot outrank
  a valid stated bound simply because it contains more points. Point count
  is the last selection criterion.
\item \emph{Reading supplied data and vector coordinates directly.} Numerical
  files and PDF vector paths allow exact or near-exact extraction
  (\ref{sec:channels}) with little model cost. For vector paths,
  the model identifies the limit curve, while code reads its coordinates.
  These methods provide a reference for assessing extraction from images.
\item \emph{Deterministic guards.} These automatic checks recorded candidate rejections or demotions on 50 of 329 papers in the Fable~5 pipeline benchmark. Instructions asking the model to be careful do not
  enforce these checks.
\item \emph{Managing conversions in one place.} Every channel identifies the
  quantity and convention it returns, and the vision channel preserves the
  plotted values. Only the reviewed conversion registry performs conversions.
  Applying a conversion twice or skipping it entirely previously turned
  correct readings into errors of about 8~dex (\ref{sec:anatomy}).
\item \emph{Requiring review for unknown conventions.} When no approved
  conversion exists, the pipeline flags the paper for human review. This
  happened for 20 papers in the production benchmark, making the uncertainty
  visible to reviewers.
\item \emph{Checking curves against values in the text.} A numerical bound
  stated in the text provides an inexpensive, independent check on a traced
  curve. Agreement supports the extraction; disagreement identifies a case
  that needs further checking.
\item \emph{Expert review before merging.} Every proposed update reaches a
  domain expert before it can enter the repository, including errors that
  automatic checks miss.
\item \emph{Checking service availability and model identity}
  (\ref{sec:ops}). The pipeline stops when the service is unavailable and
  verifies which model is running before extraction begins.
\item \emph{Separating model effects from pipeline changes and variation between
  runs} (\ref{sec:repeatability}). Both are needed to establish the cause
  of an apparent improvement and whether it exceeds ordinary run-to-run
  variation.
\item \emph{Caching prompts only when they will be reused.} A one-hour cache
  can reduce costs for repeated reads or voting ($N{>}1$). Caching each paper's
  text and images during a single read instead roughly doubled the cost,
  because those inputs were never reused.
\end{itemize}
These results support three broader principles: use the LLM to read
  and interpret papers while code handles arithmetic, conversion, and
  selection; record declarations, notes, and check results so both code and
  reviewers can inspect the model's output; and measure variation between
  runs before claiming an improvement.
\subsection{What did not help}
\begin{itemize}
\item \emph{Combining repeated extractions.} We compared consensus from three
  reads ($N=3$) with a single read ($N=1$) on 30 papers,\footnote{This test used Opus~4.8.} using the same
  transport (the method used to access the model). The sample emphasized difficult figure extractions,
  where repeated reads were expected to help most. The improvement was only
  0.003~dex, at three times the extraction cost. This small test cannot rule
  out benefits below the measured run-to-run variation
  (\ref{sec:repeatability}). Channel selection also varied between runs despite the use of consensus.
  The pipeline therefore uses one read per extraction channel.
\item \emph{Asking the model to convert units.} Allowing both the model and
  the registry to convert values sometimes led to conversions being applied twice or
  skipped entirely (\ref{sec:anatomy}). Requiring the model to preserve the
  plotted values and assigning all conversions to the registry resolved this
  ambiguity.
\item \emph{Deriving unknown conversions during extraction.} We tested
  deriving conversions during extraction and applying them provisionally,
  subject to the same automatic checks as reviewed conversions. On papers
  flagged for convention review, the checks rejected conversions that failed their criteria but
  recovered only a few cases. Most of the reduction in unsupported cases came
  from improved identification of coupling types and figure-axis quantities.
  These results favour deriving and reviewing conversions before adding them
  to the pipeline (\ref{sec:conv}). However, we do not exclude the possibility of better harnessing (which might require higher cost), or future frontier LLM with stronger physics capability can provide more reliable unknown conversion derivation during extraction.
\item \emph{Recognising conventions from partial text matches.} A familiar
  symbol within the model's declaration does not establish that the declared
  quantity has a supported conversion. For example, the product $g_s g_p$
  contains the symbol $g_p$, but represents a different quantity. Matching
  that symbol alone incorrectly allowed conversion and suppressed human
  review. The pipeline now checks the full declaration against the reviewed
  definitions.
\item \emph{Interpreting small differences from a single run.} A
  0.02--0.06~dex difference between single runs may reflect changes in channel
  selection rather than an improvement. Such differences need to be assessed
  against the measured variation between runs (\ref{sec:repeatability}).
\end{itemize}

\subsection{Pipeline accuracy by extraction channel and coupling type}
\label{sec:pipeline-channel-accuracy}
Broken down by the selected extraction channel, the Fable~5 pipeline has
median residuals of 0.00~dex for supplied numerical data ($N=3$), 0.18~dex
for tables ($N=4$), and 0.15~dex for vector traces ($N=9$). The two largest
groups, text and figure-based extraction, have median residuals of 0.18~dex
($N=129$) and 0.13~dex ($N=116$), respectively. These groups contain different
papers, so their differences do not establish that one extraction channel
is intrinsically more accurate.

The pipeline results also vary across coupling types (figure~\ref{fig:pertype}).
The pattern is not simply that rare types are harder: AxionEDM has a median
residual of 0.13~dex across three papers. Larger discrepancies occur for
AxionMass (1.01~dex, $N=12$) and ScalarNucleon (7.00~dex, $N=2$). Because the
macro average weights each coupling type equally, these small groups have a
substantial influence; five of the twelve types contain at most five papers.\footnote{The macro average is tail-sensitive by construction (it cannot be bought by accuracy on the most common type), but
with several types represented by few papers it is also noisy and
carries no measured run-to-run floor, so we use it qualitatively, to flag
tail-heavy types, rather than as a precise figure of merit. The average includes the types with finite comparisons: 13 for agents and 12 for pipelines.}

\subsection{How conversion errors produced large discrepancies}
\label{sec:anatomy}
Incorrect handling of units and variables caused errors of several orders
of magnitude even when the figure was read correctly. The following examples show how two checks
prevent large conversion errors.\footnote{The quoted example residuals are from the Opus~4.8 pipeline.}

\emph{Applying each conversion once.} One paper plots the squared electron
coupling $(g_p^e)^2$~\cite{Terrano:2015sna}. When conversion was left partly
to the model, the square root could be taken twice, once by the model and
again by the registry, or not at all if each assumed the other had done it.
Both mistakes turned correct readings into errors of about 8~dex. The model
is now required to return the plotted values and identify their quantity,
leaving the registry to convert them exactly once. The same paper then
scores 0.04~dex.

\emph{Checking that values match the declared quantity.} The model can still
return values that it has already converted, so its declaration alone is
insufficient. Each converter therefore checks whether the input magnitudes
are plausible for the declared variable. For the small couplings considered
here, squared values are far below the expected range of unsquared values.
If a curve is declared squared but its values lie in the unsquared range,
the pipeline treats the declaration as incorrect and compares the values
without taking another square root. In one example, this preserves a reading
within 0.06~dex of the curator's curve that would otherwise acquire a 6-dex
error. The three papers that trigger this check score 0.13, 0.2, and 3.4~dex;
the last still has an extraction error, but no longer a conversion error.

\section{Benchmark scoring}
\label{app:selection}
\subsection{Convention handling}
\label{app:benchmark-conventions}
The conversion registry described in \ref{sec:conv} is also used in the benchmark evaluation of both workflows. Table~\ref{tab:conventions} gives representative examples.

\paragraph{Handling unusual data files}
Some files contain very large values, such as $10^{20}$, $10^{30}$, or
$10^{99}$, used to close a shaded polygon in a plot. These are not measured
bounds and are removed using a threshold of $10^{19}$. Some scalar-coupling
files also have headers naming $d_e$ even though their numerical values are
inconsistent with that quantity. In those cases, classification uses the
value range rather than the header.

\paragraph{What the benchmark can and cannot check}
Both curves are converted to the same canonical variable before their
residual is calculated. Multiplying both by the same factor leaves their
logarithmic difference unchanged, so the benchmark cannot establish whether
that shared factor is correct. Such factors require separate tests in code
and checks against the literature. Conversions applied differently to the two
curves can affect the residual and are therefore tested by the comparison.
This usually occurs on the extracted side, where declared conventions vary
between papers. The cancellation described above applies to shared
multiplicative factors, not to conversions in general.

\paragraph{Magnitude-based convention overrides}
Using the coupling magnitude to override a declared convention is a heuristic, not an independent verification of that convention. In the finite pipeline comparisons,\footnote{These counts concern overrides in comparisons yielding a finite residual, which are the cases entering the reported residual statistics. Papers without a finite comparison do not enter those statistics. The counts do not include corrections made during extraction.} the scorer applied such overrides in 7 of 261 Fable~5 cases, 3 of 259 Opus~4.8 cases, and 4 of 222 Haiku~4.5 cases. No such override occurred for either the extracted curves or the selected references in the finite agent comparisons.

\begin{table}[htbp]
\caption{Representative examples of the per-file convention conversions the
registry applies to both the extracted and the reference curve before scoring;
this is a subset chosen to illustrate the range of cases, not the full set. The
factors were derived with GPD~\cite{gpd}, checked in code against the repository's
plotting source, and verified against the cited literature. The
``SNO file only'' row is a convention specific to a single Sudbury Neutrino
Observatory data file rather than a coupling-wide rule.
\emph{Notation.}
$f_a$ is the axion decay constant and $C_N$ the model-dependent nucleon
coupling coefficient~\cite{diluzio_landscape}; the \emph{derivative}
(pseudovector) coupling $g_{aNN}=C_N/2f_a$~[GeV$^{-1}$] is the coefficient of
the $(\partial_\mu a)\,\bar N\gamma^\mu\gamma_5 N$ interaction, which on shell
maps to the dimensionless pseudoscalar coupling $g_{aN}=2m_N\,g_{aNN}$.
$\chi=\varepsilon$ is the dark-photon kinetic mixing~\cite{dp_cookbook},
$d_e,d_{m_e}$ are dilaton couplings~\cite{damour_donoghue_dilaton}, $\alpha$ a
fifth-force strength relative to gravity, and $g_{an\gamma}\equiv g_d$ the
axion--nucleon electric-dipole coupling. The nucleon masses stored in the
registry are $m_n=0.93957$~GeV and $m_p=0.93828$~GeV (the $10^{-5}$ rounding
difference from the PDG proton mass is negligible at the reported precision).}
\label{tab:conventions}
\centering
\footnotesize
\setlength{\tabcolsep}{5pt}
\renewcommand{\arraystretch}{1.3}
\resizebox{\textwidth}{!}{%
\begin{tabular}{@{}p{2.4cm}p{3.0cm}p{2.8cm}p{5.1cm}@{}}
\toprule
\textbf{Coupling type} & \textbf{Canonical variable} & \textbf{Common paper variable} & \textbf{Conversion to canonical variable} \\
\midrule
AxionNeutron /\newline AxionProton & dimensionless $g_{aN}$ & derivative coupling $g_{aNN}$ [GeV$^{-1}$] & $g_{aN}=2m_N\,g_{aNN}$ \\
AxionNeutron (SNO file only) & dimensionless $g_{an}$ & stored $g_{an}/m_n$ [GeV$^{-1}$] & $g_{an}=m_n\times(\text{file value})$ \\
Scalar / dilaton & dimensionless $d_e,\,d_{m_e}$ & fifth-force strength $\alpha$ & $d=Q^{-1}\sqrt{\alpha}$, with $Q^{-1}=500$ (photon, nucleon) or $4000$ (electron) \\
DarkPhoton & kinetic mixing $\chi$ & $\varepsilon^2$ & $\chi=\sqrt{\varepsilon^2}$ \\
AxionEDM & $g_{an\gamma}$ [GeV$^{-2}$] & $1/f_a$ [GeV$^{-1}$] & $g_{an\gamma}=3.7\times10^{-3}\ \text{GeV}^{-1}\times(1/f_a)$ \\
AxionEDM & $g_{an\gamma}$ [GeV$^{-2}$] & oscillating-EDM amplitude $d_n$ [e$\cdot$cm] & $g_{an\gamma}=d_n\,m_a/\sqrt{2\rho_{\rm DM}}$ (mass-dependent, per point) \\
AxionMass & normalized $1/f_a$ coupling & $f_a$ [GeV] & $f_a\to 1/f_a$ per point \\
\bottomrule
\end{tabular}%
}
\end{table}

\subsection{Curve comparison and reference selection}

Two steps make the per-paper comparison more reliable. First, if no reference mass can be compared by interpolating the extracted curve, we instead interpolate the reference curve at extracted masses within its range. The residuals from the two directions are not combined; the second direction is used only when the first yields no comparable points. Second, a paper may contribute
several reference curves, for different couplings or different files of the
same result. When several eligible reference curves are available, we select
the one with the smallest finite median residual over the overlapping mass
range and use that reference for all evaluation metrics. Choosing the closest reference curve does not by itself confirm that it represents the same physical scenario as the extracted result.
Compared with choosing the eligible reference with the most usable mass points, selecting the closest reference changes the choice for 11, 8, and 4 agent papers and 4, 4, and 6 pipeline papers for Fable~5, Opus~4.8, and Haiku~4.5, respectively. These represent 1.6--4.0\% of the papers eligible for curve evaluation in each configuration.

\subsection{Label evaluation and evaluation outcomes}

Two details matter when evaluating coupling-type predictions. First, some
papers report several couplings while the repository includes only one.
Predicting any coupling actually reported in the paper counts as correct,
even if the benchmark lacks its reference curve. Second, gauged $U(1)_B$ and
$U(1)_{B\text{-}L}$ bosons can use the same dimensionless symbol
$\varepsilon$ as dark-photon kinetic mixing. A check of the declared
convention assigns these papers to the $B\!-\!L$ category, addressing the
most common classification confusion observed during development.

For the pipeline, the other labels were compared with those from a separate LLM.
This measures \emph{agreement between models}, because agreement alone does not
establish correctness. The second model was also checked against a human
annotator on 15 papers. They agreed on 15/15 labels for \emph{is new limit},
15/15 for \emph{is projection}, and 14/15 for \emph{data source}, which
identifies where the curve is best read.

Table~\ref{tab:disposition} summarizes the Fable~5 agent and pipeline outcomes for all benchmark papers.

``Reference excluded from evaluation'' denotes papers whose reference data
were excluded under the documented benchmark criteria,\footnote{Examples
include theoretical mass-prediction bands, reference curves unavailable in
the supplied paper, and documented reference-data problems. Per-reference
reasons are distributed with the benchmark~\cite{axionlimitbench_huggingface}.}
independently of the extraction result. ``No reference for the extracted
coupling'' means that the reference collection contains no curve for the
coupling returned by the extraction; this can reflect either a
misclassification or a coupling genuinely reported in the paper but absent
from the collection.

\begin{table}[htbp]
\caption{Evaluation outcomes for all 329 benchmark papers using Fable~5 with
agent and pipeline extraction. Finite comparisons enter the residual
statistics; the additional exclusions for coverage evaluation are described
in section~\ref{sec:metrics}.}
\label{tab:disposition}
\centering
\small
\begin{tabular}{@{}lrr@{}}
\toprule
\textbf{Evaluation outcome} & \textbf{Agent} & \textbf{Pipeline} \\
\midrule
Finite numerical comparison & 281 & 261 \\
No comparable mass points & 0 & 4 \\
No reference for the extracted coupling & 7 & 14 \\
Unresolved convention & 0 & 9 \\
No extracted points & 1 & 2 \\
No usable reference data & 2 & 1 \\
Reference excluded from evaluation & 38 & 38 \\
\midrule
\textbf{Total} & \textbf{329} & \textbf{329} \\
\bottomrule
\end{tabular}
\end{table}

\section{Cost and model choice}
\label{sec:cost}
AxionLimitBench reports median agent extraction costs of \$1.48 per paper for Fable~5, \$0.99 for Opus~4.8, and \$0.28 for Haiku~4.5~\cite{yuan2026axionlimitbench}. These are nominal API-equivalent charges reported by the command-line interface. We adopt the Fable~5 agent for its higher joint success rate (table~\ref{tab:results}).

For comparison, the pipeline's list-price estimates are \$0.88, \$0.44, and \$0.09 per paper for the same three models, respectively~\cite{yuan2026axionlimitbench}. These computational costs exclude human review and correction, and any reduction in expert time remains unmeasured.

\section{Coverage, confidence, and repeatability}
\label{sec:calib}
\begin{figure}[htbp]
\centering
\includegraphics[width=0.95\textwidth]{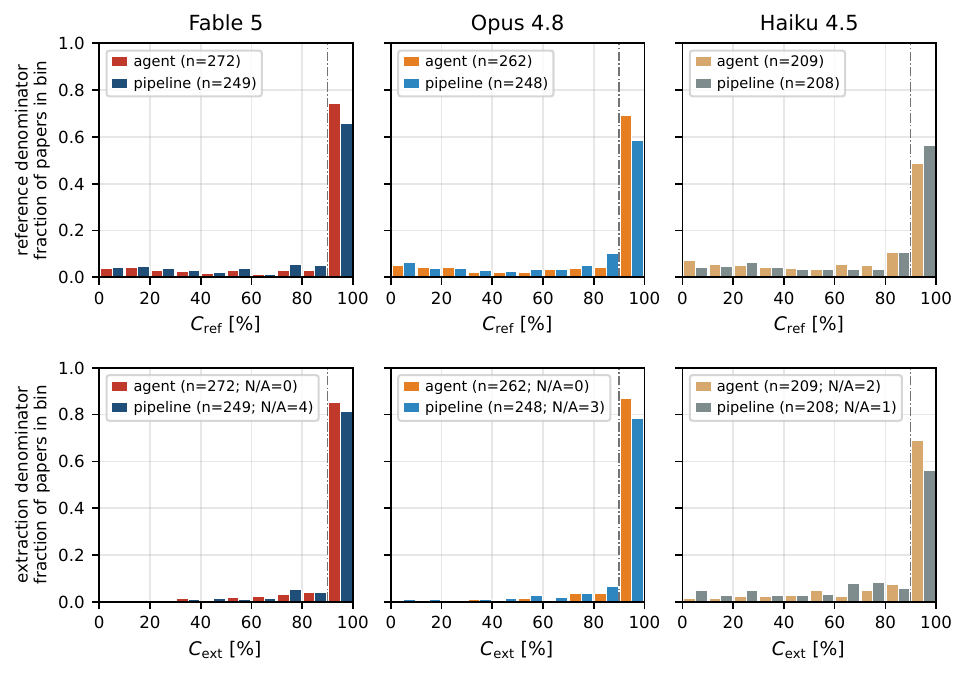}
\caption{Distributions of reference coverage (top) and extraction coverage
(bottom), measured over logarithmic mass ranges as defined in
section~\ref{sec:metrics}. Bars show the fraction of eligible papers in each
10-percentage-point bin; vertical lines mark 90\% coverage. Sample sizes are
given in the legends. ``N/A'' denotes point-only extractions, for which
extraction coverage is undefined; these remain in the denominator and fail
the coverage criterion.}
\label{fig:coverage-distributions}
\end{figure}

\paragraph{Coverage distributions}
Figure~\ref{fig:coverage-distributions} shows the distributions of both
coverage measures for all six configurations.

\begin{figure}[htbp]
\centering
\includegraphics[width=0.9\textwidth]{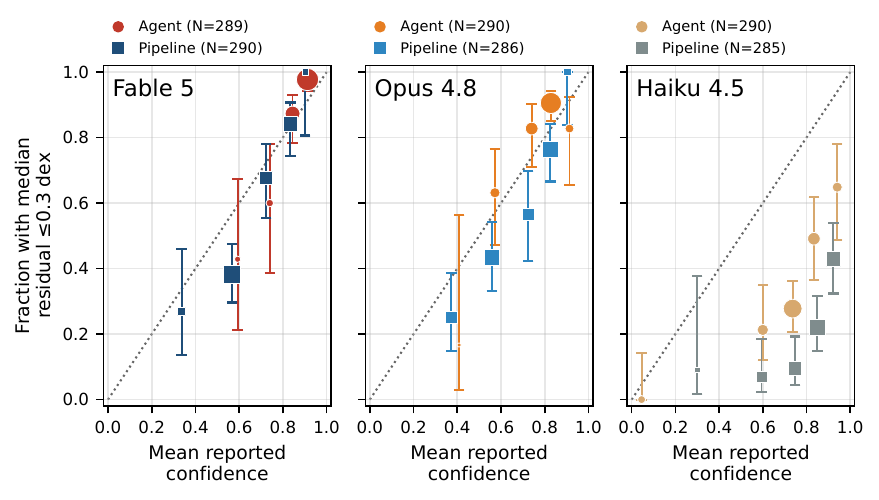}
\caption{Confidence and measured accuracy for the six benchmark configurations.
Papers are grouped into fixed confidence bins. The $y$-axis gives the fraction
in each group meeting the calibration success criterion: median residual
$\leq0.3$~dex. Marker area is proportional to group size; error bars are 95\%
Wilson confidence intervals. Low confidence flags results for closer human
review; all proposals require review.}
\label{fig:calib}
\end{figure}

\paragraph{Confidence and extraction accuracy}
To assess whether higher confidence corresponds to better extractions, we
group papers by the model's reported confidence (figure~\ref{fig:calib}).
We use the same fixed confidence bins for all six configurations and measure
the fraction of papers with a median residual $\leq0.3$~dex.\footnote{Only papers with a reported confidence and without excluded or unusable reference data are included. Other cases lacking a finite residual count as failures.} This diagnostic
does not require sufficient coverage and is therefore distinct from the
joint success criterion in table~\ref{tab:results}. Higher confidence generally
corresponds to better extraction accuracy, but the relationship varies across
models and workflows; confidence should not be interpreted directly as a
probability of success.

The pipeline uses confidence to flag extractions for closer review and to
break remaining ties in candidate selection. All proposed repository updates
require expert review.

\begin{figure}[htbp]
\centering
\includegraphics[width=0.95\textwidth]{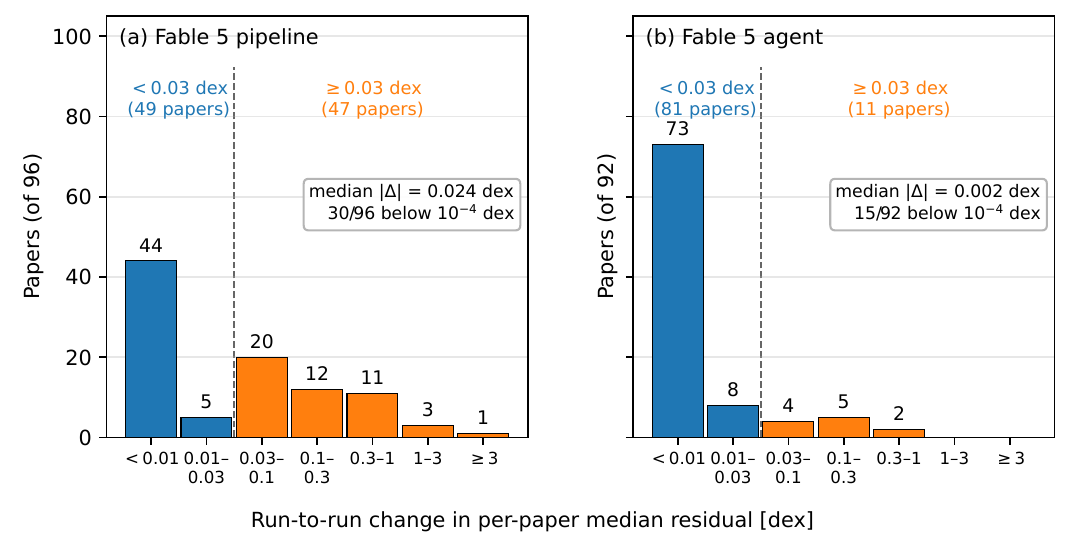}
\caption{Per-paper absolute change in the median residual between two
Fable~5 runs, for the pipeline (left; 96 paired papers) and agent (right;
92 paired papers). Each panel uses a frozen random 100-paper subset: the
pipeline compares its benchmark run with a fresh read, while the agent
compares two fresh sessions with the same configuration. Blue bars show
changes below 0.03~dex (49 pipeline papers and 81 agent papers); orange bars
show larger changes.}
\label{fig:noisefloor}
\end{figure}

\paragraph{Variation between repeated runs}
\label{sec:repeatability}
For the pipeline, three runs on the same fixed subset of 100 papers give a
median per-paper standard deviation of 0.04~dex among the 93 papers with
finite residuals in all three runs. Between pairs of runs,
12\% of papers change extraction channel; 28\% change at
least once across all three runs. These switches produce some large changes
in the per-paper median residual, reaching approximately 7~dex, while half
the papers differ by less than 0.03~dex between a pair of runs
(figure~\ref{fig:noisefloor}). For the previous production model, Opus~4.8,
one repeat read of 56 papers gives an 18\% channel-switching rate.
On these same 56 papers, the bootstrap 95\% interval for the aggregate
median residual has a half-width of 0.05~dex for Opus~4.8 and 0.04~dex for
Fable~5.\footnote{Each bootstrap sample randomly selects one of the two
available residuals for each paper and recomputes the median. These intervals
describe variation under this resampling procedure, rather than the spread
of complete repeated runs.}

For the agent, 100 papers were drawn uniformly at random from the benchmark
manifest's 292-paper sampling pool\footnote{This sampling list includes one
paper whose reference is excluded by the scorer used for the main results.
That paper was not among the 100 selected.} and processed twice with the
same configuration.
The runs yielded finite residuals for 92 and 95 papers, respectively, with
92 in common; three papers were comparable only in the second run and five
in neither. The selected reference curve remained unchanged for all 92
paired papers. Direct comparison of the extracted curves over their
overlapping mass range gives a median per-paper disagreement of 0.005~dex,
with a 90th percentile of 0.127~dex. Three papers changed coupling plane
between runs, and one first-run extraction contained no points; these
outcomes remain included in the reported totals.

\newpage
\section{Agent task prompt}
\label{app:agent-prompt}
We reproduce below the task prompt specifying the extraction objective, permitted sources, coupling conventions, and output format.
\begingroup
\small
\subsection*{AxionLimitBench task card (v0.1.1)}

You are extracting an experimental exclusion limit from a particle-physics paper. The output is judged against the curve that the maintainer of the AxionLimits compilation digitised for the same paper, so the goal is the curve a careful physicist would add to that compilation: the paper's headline measured limit, in the canonical variable and units listed below, as a list of (mass, coupling) points.

\subsubsection*{Inputs you may use}

\begin{itemize}
\item The paper itself: \texttt{./paper.pdf} in the current working directory.
\item The paper's own arXiv e-print (LaTeX source, embedded figure files, ancillary data files): \url{https://arxiv.org/e-print/<id>} or \url{https://arxiv.org/src/<id>}.
\item Any code you write yourself. \texttt{python3} is available with numpy, scipy, pymupdf (\texttt{fitz}), matplotlib, PIL, cv2 and skimage.
\end{itemize}

You must NOT consult secondary sources: no limit compilations (AxionLimits, cajohare, DarkCast), no HEPData, INSPIRE, PDG, no other papers, no web search. Everything you report must come from this paper and its e-print. Sessions are audited for this.

\subsubsection*{What to decide}

\begin{enumerate}
\item \textbf{Is there a new measured limit?} \texttt{is\_\allowbreak{}new\_\allowbreak{}limit} is true only if the paper reports its own experimental or observational exclusion on one of the couplings below. Projections, sensitivity forecasts and proposals are \texttt{is\_\allowbreak{}projection: true} and are out of scope (report them as such with no points). Pure theory papers and reinterpretations of other experiments' data are not new limits unless the paper derives a new bound itself.
\item \textbf{Which coupling?} One \texttt{coupling\_\allowbreak{}type} from the table. If the paper bounds several, pick the one it presents as its main result.
\item \textbf{The curve.} The boundary of the excluded region, as points \texttt{[mass\_\allowbreak{}eV, coupling]} with the coupling in the canonical variable and units of the table, ordered by mass, positive finite numbers only, typically 10 to 100 points (fewer if the limit is a flat line or a single value; a flat limit over a stated mass range is two points). Trace the paper's headline curve (the observed limit, not the expected sensitivity band).
\item \textbf{Metadata}, as in the schema below.
\end{enumerate}

\subsubsection*{Canonical planes}

x-axis is always the boson mass in eV. y-axis:

\begingroup\scriptsize\setlength{\tabcolsep}{3pt}\renewcommand{\arraystretch}{1.1}
\begin{longtable}{@{}>{\raggedright\arraybackslash}p{0.145\linewidth}>{\raggedright\arraybackslash}p{0.15\linewidth}>{\raggedright\arraybackslash}p{0.14\linewidth}>{\raggedright\arraybackslash}p{0.305\linewidth}>{\raggedright\arraybackslash}p{\dimexpr0.26\linewidth-24pt\relax}@{}}
\toprule
coupling\_\allowbreak{}type & variable & units & definition & exact \texttt{coupling\_\allowbreak{}convention} label \\ \midrule
\endfirsthead
\toprule
coupling\_\allowbreak{}type & variable & units & definition & exact \texttt{coupling\_\allowbreak{}convention} label \\ \midrule
\endhead
\bottomrule\endfoot
Axion\allowbreak{}Photon & g\_\allowbreak{}a\ensuremath{\gamma}\ensuremath{\gamma} & GeV\textasciicircum{}-1 & L = -(g\_\allowbreak{}a\ensuremath{\gamma}\ensuremath{\gamma}/4) a F \ensuremath{\widetilde F} & \texttt{g\_\allowbreak{}agamma [GeV\textasciicircum{}-1]} \\[1pt]
Axion\allowbreak{}Electron & g\_\allowbreak{}ae & dimensionless & L = g\_\allowbreak{}ae (\ensuremath{\partial}\_\allowbreak{}\ensuremath{\mu} a / 2m\_\allowbreak{}e) \ensuremath{\bar e} \ensuremath{\gamma}\textasciicircum{}\ensuremath{\mu} \ensuremath{\gamma}\_\allowbreak{}5 e, i.e. g\_\allowbreak{}ae = C\_\allowbreak{}e m\_\allowbreak{}e / f\_\allowbreak{}a & \texttt{g\_\allowbreak{}ae (dimensionless)} \\[1pt]
Axion\allowbreak{}Neutron & g\_\allowbreak{}an & dimensionless & L = (g\_\allowbreak{}an / 2m\_\allowbreak{}n) \ensuremath{\partial}\_\allowbreak{}\ensuremath{\mu} a \ensuremath{\bar n} \ensuremath{\gamma}\textasciicircum{}\ensuremath{\mu} \ensuremath{\gamma}\_\allowbreak{}5 n, i.e. g\_\allowbreak{}an = C\_\allowbreak{}n m\_\allowbreak{}n / f\_\allowbreak{}a; \textbf{see the 2m\_\allowbreak{}N rule below} & \texttt{g\_\allowbreak{}an (dimensionless)} \\[1pt]
Axion\allowbreak{}Proton & g\_\allowbreak{}ap & dimensionless & same normalisation with the proton; \textbf{see the 2m\_\allowbreak{}N rule below} & \texttt{g\_\allowbreak{}ap (dimensionless)} \\[1pt]
Axion\allowbreak{}EDM & g\_\allowbreak{}d (also written g\_\allowbreak{}aN\ensuremath{\gamma}) & GeV\textasciicircum{}-2 & L = -(i/2) g\_\allowbreak{}d a \ensuremath{\bar N} \ensuremath{\sigma}\_\allowbreak{}\ensuremath{\mu}\ensuremath{\nu} \ensuremath{\gamma}\_\allowbreak{}5 N F\textasciicircum{}\ensuremath{\mu}\ensuremath{\nu}; the oscillating nucleon EDM is d\_\allowbreak{}n = g\_\allowbreak{}d a & \texttt{g\_\allowbreak{}d [GeV\textasciicircum{}-2]} \\[1pt]
Axion\allowbreak{}Mass & 1/f\_\allowbreak{}a & GeV\textasciicircum{}-1 & inverse axion decay constant in the convention where the QCD axion has m\_\allowbreak{}a = 5.7 \ensuremath{\mu}eV \ensuremath{\times} (10\textasciicircum{}12 GeV / f\_\allowbreak{}a); used for bounds quoted directly on f\_\allowbreak{}a, e.g. CP-violating or EDM-type searches & \texttt{1/f\_\allowbreak{}a [GeV\textasciicircum{}-1]} \\[1pt]
Dark\allowbreak{}Photon & \ensuremath{\chi} (kinetic mixing, also \ensuremath{\epsilon}) & dimensionless & L \ensuremath{\supset} -(\ensuremath{\chi}/2) F\_\allowbreak{}\ensuremath{\mu}\ensuremath{\nu} F'\textasciicircum{}\ensuremath{\mu}\ensuremath{\nu}; report \ensuremath{\chi}, never \ensuremath{\chi}\textsuperscript{2} & \texttt{kinetic mixing chi (dimensionless)} \\[1pt]
Vector\allowbreak{}BL & g\_\allowbreak{}B-L & dimensionless & gauge coupling of the U(1)\_\allowbreak{}B-L boson; report g, never \ensuremath{\alpha} = g\textsuperscript{2}/4\ensuremath{\pi} & \texttt{g\_\allowbreak{}BL (dimensionless)} \\[1pt]
Scalar\allowbreak{}Photon & d\_\allowbreak{}e & dimensionless & Damour--Donoghue dilatonic coupling, \ensuremath{\alpha} \ensuremath{\to} \ensuremath{\alpha}(1 + d\_\allowbreak{}e \ensuremath{\kappa} \ensuremath{\phi}), \ensuremath{\kappa} = \ensuremath{\surd}(4\ensuremath{\pi})/M\_\allowbreak{}Pl, M\_\allowbreak{}Pl = 1.22\ensuremath{\times}10\textasciicircum{}19 GeV & \texttt{d\_\allowbreak{}e (dimensionless)} \\[1pt]
Scalar\allowbreak{}Electron & d\_\allowbreak{}me & dimensionless & m\_\allowbreak{}e \ensuremath{\to} m\_\allowbreak{}e(1 + d\_\allowbreak{}me \ensuremath{\kappa} \ensuremath{\phi}), same \ensuremath{\kappa} & \texttt{d\_\allowbreak{}me (dimensionless)} \\[1pt]
Scalar\allowbreak{}Nucleon & g\_\allowbreak{}N (or the paper's own dimensionless scalar--nucleon coupling) & dimensionless & L = g\_\allowbreak{}N \ensuremath{\phi} \ensuremath{\bar N}N; fifth-force papers instead quote |\ensuremath{\alpha}| (Yukawa strength relative to gravity) versus range \ensuremath{\lambda}, with m = \ensuremath{\hbar}c/\ensuremath{\lambda} = 1.97\ensuremath{\times}10\textasciicircum{}-7 eV\ensuremath{\cdot}m / \ensuremath{\lambda} & \texttt{coupling (dimensionless)} \\[1pt]
Scalar\allowbreak{}Baryon & g\_\allowbreak{}B & dimensionless & coupling of a scalar to baryon number (equivalence-principle and fifth-force tests) & \texttt{coupling (dimensionless)} \\[1pt]
Monopole\allowbreak{}Dipole & g\_\allowbreak{}s g\_\allowbreak{}p & dimensionless & product of scalar and pseudoscalar couplings for a monopole--dipole spin-dependent force (electron--nucleon or nucleon--nucleon), versus mediator mass & \texttt{g\_\allowbreak{}s g\_\allowbreak{}p (dimensionless)} \\[1pt]
Axion\allowbreak{}CPV & as the paper defines & dimensionless & other CP-violating axion couplings & \texttt{coupling (dimensionless)} \\[1pt]
\end{longtable}
\endgroup

Conventions of the compilation:

\begin{itemize}
\item \textbf{Dark-matter density: do NOT rescale.} Report dark-matter-search limits (haloscopes and other direct searches) exactly as the paper states them, at the paper's own assumed local density, and record that density in \texttt{dm\_\allowbreak{}density\_\allowbreak{}assumed} (null if the limit does not depend on it). The compilation stores paper-native values and applies any density rescaling only when it draws the plot.
\item \textbf{Axion--nucleon normalisation (the 2m\_\allowbreak{}N rule).} The canonical g\_\allowbreak{}an and g\_\allowbreak{}ap are the dimensionless couplings of L = (g\_\allowbreak{}aN / 2m\_\allowbreak{}N) \ensuremath{\partial}\_\allowbreak{}\ensuremath{\mu} a \ensuremath{\bar N} \ensuremath{\gamma}\textasciicircum{}\ensuremath{\mu} \ensuremath{\gamma}\_\allowbreak{}5 N. Many papers instead write L = g \ensuremath{\partial}\_\allowbreak{}\ensuremath{\mu} a \ensuremath{\bar N} \ensuremath{\gamma}\textasciicircum{}\ensuremath{\mu} \ensuremath{\gamma}\_\allowbreak{}5 N with g in GeV\textasciicircum{}-1, or the non-relativistic Hamiltonian H = g \ensuremath{\nabla}a \ensuremath{\cdot} \ensuremath{\sigma}\_\allowbreak{}N; those are the same physics with g\_\allowbreak{}aN = 2 m\_\allowbreak{}N g, \textbf{not} m\_\allowbreak{}N g. Example: g = 1.0\ensuremath{\times}10\textasciicircum{}-5 GeV\textasciicircum{}-1 gives g\_\allowbreak{}an = 2 \ensuremath{\times} 0.939 GeV \ensuremath{\times} 1.0\ensuremath{\times}10\textasciicircum{}-5 GeV\textasciicircum{}-1 = 1.9\ensuremath{\times}10\textasciicircum{}-5. If the paper bounds a combination such as the isovector |g\_\allowbreak{}an \ensuremath{-} g\_\allowbreak{}ap|/2, report the bound on the single canonical coupling with the other set to zero and say so in \texttt{notes}.
\item \textbf{Scenarios.} If the paper shows several curves (e.g. polarisation scenarios, model variants, prior choices), report the one it presents as its main result and say which in \texttt{notes}.
\item \textbf{Declare what you emitted.} \texttt{coupling\_\allowbreak{}convention} must be EXACTLY the label from the last column of the table for the coupling you report, e.g. \texttt{"g\_\allowbreak{}d [GeV\textasciicircum{}-2]"}. It is parsed by a strict matcher: do not add symbols, definitions, formulas or the paper's own notation to it (a declaration like \texttt{"g\_\allowbreak{}adgamma [GeV\textasciicircum{}-2] (d\_\allowbreak{}d = g a)"} is rejected as unknown even when the numbers are right). Put derivations, equation numbers and conversions in \texttt{notes}. Only if you genuinely could not convert to the canonical variable, emit the paper's native variable and declare it with one of these native labels instead: \texttt{"d\_\allowbreak{}n [e cm]"} (oscillating nucleon EDM amplitude), \texttt{"f\_\allowbreak{}a [GeV]"}, \texttt{"chi\textasciicircum{}2 (dimensionless)"}, \texttt{"alpha (Yukawa strength relative to gravity)"}. A truthful native declaration is far better than a guessed conversion.
\end{itemize}

\newpage
\subsubsection*{Output: \texttt{./result.json}}

\begin{Verbatim}[fontsize=\footnotesize,breaklines=true,breakanywhere=true]
{
  "paper_title": "string",
  "coupling_type": "AxionPhoton | AxionElectron | ... | null",
  "is_new_limit": true,
  "is_projection": false,
  "data_points": [[1.0e-6, 3.2e-14], [2.0e-6, 2.9e-14]],
  "data_source": "table | text | figure_vision | figure_vector | ancillary_file | none",
  "dm_density_assumed": 0.45,
  "confidence_level": 0.95,
  "coupling_convention": "g_agamma [GeV^-1]",
  "suggested_experiment_name": "ShortName_Year",
  "extraction_confidence": 0.8,
  "notes": "which figure/table/equation the curve came from, scenario chosen, conversions applied, caveats"
}
\end{Verbatim}

\begin{itemize}
\item \texttt{data\_\allowbreak{}source}: where the numbers came from (\texttt{ancillary\_\allowbreak{}file} for a data file shipped with the e-print, \texttt{figure\_\allowbreak{}vector} for coordinates recovered from a vector figure, \texttt{figure\_\allowbreak{}vision} for reading a raster/vector figure by eye, \texttt{table} / \texttt{text} for tabulated or quoted values).
\item \texttt{dm\_\allowbreak{}density\_\allowbreak{}assumed}: the density the paper assumed in GeV/cm\textsuperscript{3}, or null when the limit does not depend on it.
\item \texttt{confidence\_\allowbreak{}level}: 0.90 or 0.95 as the paper states.
\item \texttt{extraction\_\allowbreak{}confidence}: your calibrated probability (0 to 1) that the curve is within a factor of two of the paper's headline limit over its mass range.
\item If there is no measured limit to report, still write the file with \texttt{is\_\allowbreak{}new\_\allowbreak{}limit: false} (or \texttt{is\_\allowbreak{}projection: true}) and \texttt{data\_\allowbreak{}points: []}.
\end{itemize}

Write the file, then stop. Do not print the points in your final reply.

\endgroup

\section*{Data and software availability}
The AutoAxionLimits source code is available at
\url{https://github.com/FaroutYLq/AutoAxionLimits}, with the original pipeline
release archived on Zenodo~\cite{autoaxionlimits_v020}.
The benchmark reference data, frozen scorer, extraction prompts, agent driver,
and saved predictions are distributed through
\url{https://github.com/FaroutYLq/AxionLimitBench}; full agent transcripts are
available in its Hugging Face dataset~\cite{axionlimitbench_huggingface}.
The revised coverage analyses, figure-generation code, and 100-paper agent
repeatability study are included in the versioned reproducibility
supplement~\cite{axionlimitbench_pdu_supplement}.
The upstream constraint data are from the \texttt{AxionLimits}
repository~\cite{axionlimits}.
The archived outputs can be analysed without model access; repeating extraction
requires the recorded model versions and access configuration.

\section*{Funding}
Work at Washington University in St. Louis was supported in part by a McDonnell Center for the Space Sciences (MCSS) seed grant (PI: K.R.)

This work was also supported by Dianzhuo Wang and TwentyTwo. We thank Anthropic's AI for Science program for providing Claude Code credits used in this research. The funders had no
role in the study design; collection, analysis, or interpretation of data;
writing of the manuscript; or decision to submit it for publication.

\newpage
\section*{CRediT authorship contribution statement}
\textbf{Lanqing Yuan:} Conceptualization, Methodology, Software, Validation,
Formal analysis, Investigation, Data curation, Visualization, Funding acquisition,
Project administration, Writing -- original draft, Writing -- review and editing.
\textbf{Karthik Ramanathan:} Supervision, Formal analysis, Writing -- review and editing.

\section*{Declaration of competing interest}
Lanqing Yuan reports financial support from Dianzhuo Wang and TwentyTwo.
The authors declare no other known competing financial interests or personal
relationships that could have appeared to influence the work reported in this
paper.

\section*{Declaration of generative AI and AI-assisted technologies in the manuscript preparation process}
Under human direction, Claude models (through Claude Code, based on
Claude Opus~4.8) assisted with writing and editing the pipeline code and
with drafting and editing this manuscript. OpenAI Codex assisted with
revising the manuscript's language, formatting, and references. The authors
reviewed and edited the resulting content and take full responsibility for
all scientific claims, numerical results, figures, references, and the final text.

AI tools also form part of the research methods described in this paper.
The pipeline's extraction and review agents use Anthropic Claude models:
\texttt{claude-fable-5} in the production configuration, with
\texttt{claude-opus-4-8} and \texttt{claude-haiku-4-5} as the comparison arms
(section~\ref{sec:results}). The convention-registry conversion factors were
derived and checked against the cited literature with assistance from Get
Physics Done (GPD), developed by Physical Superintelligence PBC
(PSI)~\cite{gpd}, and then code-verified by the authors.

\section*{Acknowledgements}
We thank Ciaran O'Hare, whose creation and years-long maintenance of the
community \texttt{AxionLimits} repository~\cite{axionlimits} provide both the
foundation this work builds on and the curated ground truth it is measured
against.

We thank Yue Ma and Aobo Li for useful discussions.

Lanqing Yuan acknowledges OpenAI for providing access to ChatGPT and Codex through the ChatGPT for Academic Researchers program.

\bibliographystyle{elsarticle-num}
\bibliography{refs}

\end{document}